\def\re#1{(\ref{#1})}
\def\beq{\begin{equation}}
\def\eeq{\end{equation}}
\def\beeq{\begin{eqnarray}}
\def\beeqn{\begin{eqnarray*}}
\def\eeeq{\end{eqnarray}}
\def\eeeqn{\end{eqnarray*}}

                  \def\G{\Gamma}
\def\de{\delta}                 \def\D{\Delta}

\def\l{\lambda}                 
\def\m{\mu}
\def\n{\nu}

\def\s{\sigma}                  
\def\th{\theta}

\def\z{\zeta} 

\newcommand{\NN}{{\cal N}}
\newcommand{\OO}{{\cal O}}

\newcommand{\lp}{\left(}
\newcommand{\rp}{\right)}
\renewcommand{\lq}{\left[}
\renewcommand{\rq}{\right]}

\newcommand{\no}{\nonumber}
\newcommand{\ph}{\phantom} 

\def\tr{\,\mbox{Tr}\,}
\def\frac#1#2{ {{#1} \over {#2} }}

\def\half{\mbox{\small $\frac{1}{2}$}}
\def\p{\partial}

\def\ie{\hbox{\it i.e.}{ }}

\documentclass[letterpaper,notoc,12pt]{JHEP3} 

\usepackage{epsfig,multicol}
\newcommand\fverb{\setbox\pippobox=\hbox\bgroup\verb}
\newcommand\fverbdo{\egroup\medskip\noindent%
			\fbox{\unhbox\pippobox}\ }
\newcommand\fverbit{\egroup\item[\fbox{\unhbox\pippobox}]}
\newbox\pippobox

\title{Wilson line correlators in two-dimensional noncommutative
Yang-Mills theory}
\author{A. Bassetto\thanks{Partially supported by the European Community 
network HPRN-CT-2000-00149.}~ and
F. Vian\\
Dipartimento di Fisica ``G.Galilei", Via Marzolo 8, 35131
Padova, Italy\\
INFN, Sezione di Padova, Italy\\
E-mail: \email{bassetto@pd.infn.it}, \email{vian@pd.infn.it}}
\received{} 		
\revised{}
\accepted{}		

\preprint{DFPD 02/TH 19}	

\abstract{We study the correlator of two parallel Wilson lines
in two-dimensional noncommutative
Yang-Mills theory, following
two different approaches. We first consider a perturbative
expansion in the large-$N$ limit and  resum all planar diagrams.
The second approach is non-perturbative: we exploit the Morita
equivalence, mapping the 
two open lines on the noncommutative torus (which eventually gets decompacted)
in two closed Wilson loops winding around the dual commutative torus.
Planarity allows us to single out a suitable region of the variables
involved, where a saddle-point approximation
of the general Morita expression for the correlator can be performed. 
In this region  the correlator nicely compares
with the perturbative result, exhibiting an exponential increase with respect
to the momentum $p$.}

	
\dedicated{}

\begin{document} 

\section{Introduction}

Correlation functions  of gauge invariant local operators in commutative theories
provide a lot of information about the dynamics; in the presence of a 
noncommutative geometry they are  even more interesting in view
of the intimate merging of space-time and ``internal'' symmetries
\cite{Alvar,Harvey}.

The simplest way of turning ordinary theories into noncommutative ones is 
to replace  the 
usual multiplication of fields in the Lagrangian with the Moyal
$\star$-product. 
This product is realized by means of a real antisymmetric matrix 
$\theta^{\mu\nu}$ which parameterizes  noncommutativity of Minkowski 
space-time:
\beq
\label{alge}
[x^\mu,x^\nu]=i\theta^{\mu\nu}\quad\quad\quad\quad \mu,\nu=0,..,D-1.
\eeq
The $\star$-product of two fields $\phi_1(x)$ and $\phi_2(x)$ is defined as
\beq
\label{star}
\phi_1(x)\star\phi_2(x)=\exp \lq \frac{i}2 \, \theta^{\mu\nu}\frac{\partial}
{\partial x_1^\mu}
\frac{\partial}{\partial x_2^\nu}\rq \phi_1(x_1)\phi_2(x_2)|_{x_1=x_2=x}
\eeq
and  leads to terms in the action with an infinite number of derivatives of 
fields which make the theory intrinsically non-local. 
As a consequence, gauge invariance in this case is obtained only
after integration over space-time variables and the
possibility of having {\it local} probes is lost.  

It is therefore remarkable that the authors in \cite{IIKK}-\cite{Gross} have
succeeded  
in proposing a recipe in noncommutative gauge theories which turns
local operators into gauge invariant observables carrying a non-vanishing
momentum. They showed that an {\it open} Wilson line with momentum
$p_{\mu}$ is gauge invariant provided the length of the line $l^{\nu}$
is related to the momentum as follows
\beq\label{length}
l^{\nu}=p_{\mu}\theta^{\mu\nu}.
\eeq
Averaging any usual local gauge invariant operator with respect to space-time
and group variables with a weight given by an open Wilson line
provides an (over-complete) set of dynamical observables in the noncommutative
case. 
  
In \cite{Gross} a perturbative calculation of open Wilson line
correlators for a ${\cal N}=4$ supersymmetric theory in four dimensions
was carried out and compared with dual supergravity
results. Ladder diagrams dominate and a good agreement was reached.  

In two dimensions the situation is quite different: noncommutativity involves
the time variable, but the Lorentz symmetry is not violated owing to the
tensorial character of $\theta^{\mu\nu}$.  
If we choose the light-cone gauge, the perturbative calculation is greatly 
simplified, thanks to the decoupling of Faddeev-Popov ghosts
and to the vanishing of the vector vertices. 
It turns out that in all  diagrams 
contributing to the line correlators, which are planar according to
the 't Hooft's large-$N$ limit \cite{'tHooft}, 
$\th$-dependent phases resulting from non commutativity play no role.

On the other hand, in two dimensions a remarkable symmetry, the Morita
equivalence, allows the mapping of open Wilson lines on a noncommutative
torus to closed Wilson loops winding on a dual commutative torus
\cite{Schwarz}.   
In a commutative setting Wilson loop correlations can be obtained by
geometrical techniques \cite{Migdal-Rusakov}; this opens the possibility of confronting perturbative
calculations with non-perturbative solutions, provided a common kinematical region
of validity is found for both approaches.

In \cite{Mandal} three basic different regimes have been presented
for a noncommutative theory in two dimensions, when approximated
by means of a $M\times M$ matrix model. Three different phases (disordered,
planar and GMS \cite{Gopakumar}) are found according to the behaviour of the
noncommutativity parameter $\theta$ with respect to the integer $M$ which is to be sent 
eventually to $\infty$ ($\theta \sim M^{\nu}$ with $\nu <1$, $\n=1$, $\n>1$, 
respectively). This integer is in turn related to a large distance cutoff 
of the theory. Owing to the merging of space-time and ``internal'' symmetries
in a noncommutative setting \cite{Alvar,Harvey}, a large-$N$ limit forcefully entails
large $M$ and the regimes where a perturbative calculation
is likely to make sense are those in which $\theta$ is not going to increase faster
than $N$ (the disordered and the planar ones). 

Such a perturbative calculation for the correlator of two open Wilson 
lines is carried out in Sect.~3 in the large-$N$ limit. An exponential
increase with the line momentum is found, 
in agreement with an analogous result in four dimensions \cite{Gross}. In Sect.~4
the Morita equivalence is exploited together with the geometrical solutions
for the partition function and for the loop correlator
on the commutative torus. In Sect.~5 the non-perturbative result is
considered in the relevant large-$N$ region where a saddle-point approximation 
provides us with a fairly simple concrete expression for the correlator.
In this region, an exponential increase in the momentum is found
in agreement with the perturbative result. 
Outside this region, the saddle-point approximation breaks up, which
could be a signal 
of a transition to a new phase, totally invisible in the perturbative approach.
Final comments and remarks are presented in the Conclusions.

\section{Observables in noncommutative gauge theories}

To start with, we set the notation we will use throughout the paper.
We will consider the $U(N)$ Yang-Mills theory on a noncommutative
two-dimensional 
space, with classical action
\beq 
\label{act}
S=-\frac12 \int d^2x\, \tr F_{\m\n} \star  F^{\m\n}
\eeq
where the field strength $F_{\m\n}$ is given by
\beq
F_{\m\n}=\p_\m A_\n -\p_\n A_\m -ig (A_\m\star A_\n - A_\n\star A_\m)
\eeq
and $A_\m=A_\m^A t^A$ is a $N\times N$ matrix, normalized as follows: $\tr t^A
t^B=\half \de^{AB}$, capital letters denoting $U(N)$ indices.
The $\star$-product was defined in  Eq.~\re{star}.
The action    Eq.~\re{act} is invariant under $U(N)$
noncommutative gauge transformations 
\beq
\label{gauge}
\de_\l A_\m= \p_\m \l -i g (A_\m\star\l -\l\star A_\m) \,.
\eeq
As noticed in \cite{Gross}, under the transformation Eq.~\re{gauge}
the operator $\tr F^2(x)$ is not left invariant 
\beq\label{opera}
\tr F^2(x) \longrightarrow \tr U(x) \star  F^2(x) \star U^\dagger (x)\,,
\eeq
with $U(x)=\exp (ig \l (x))$. To recover a gauge invariant operator,
one has to integrate over all space, since integrals of
$\star$-product  can be cyclically permuted.
In the same fashion a noncommutative Wilson line can be  defined by
means of the Moyal product as \cite{Alvar,Gross}  
\beq
\label{wline}
\Omega_\star[x,C]=P_{\star} \exp \lp
ig\int_0^l A_\m 
(x+\z(s))\, d\z^\m(s)\rp \,,
\eeq
where $C$ is the curve 
parameterized by $\z(s)$, with $0 \leq s \leq 1$, $\z(0)=0$ and
$\z(1)=l$, and $P_\star$
denotes noncommutative path ordering along $\z(s)$ from right to left
with respect to increasing $s$ of $\star$-products of functions.
The Wilson line is not invariant under a gauge transformation
\beq\label{linetrans}
\Omega_\star [x, C] \longrightarrow  U(x) \star \Omega_\star [x,C] \star U^\dagger (x+l)\,,
\eeq
and moreover cannot be made invariant by closing the line. 
In fact, recalling that gauge invariance requires integration over
coordinates, the following operator
\beq \label{linedef0}
W(p, C)= \int d^2x \, \tr \Omega_\star[x, C]\star e^{ipx}\,,
\eeq
turns out to be invariant provided $C$ satisfies the condition
\beq\label{endpoints}
l^\n=p_\m \th^{\m\n}
\eeq
(the Wilson line extends in the direction transverse to the momentum).
For simplicity in the following only straight lines will be considered.

It is also easy to see that any local operator $\OO (x)$ in ordinary
gauge theories admits a noncommutative generalization
\beq \label{ope}
{\tilde \OO (p)}= 
\tr \int d^2x \, \OO (x) \star P_{\star}e^{ig\int_0^l d\z A(x+\z)}\star
e^{ipx}\,,
\eeq
each of the ${\tilde \OO (p)}$'s being a genuinely different operator
at different momentum.
Remarkably, at large values of $|p|$, noncommutativity imposes that
the length of the Wilson line  becomes large and the
operator is dominated by the line it is attached to. The correlation
function of these operators is thus expected to exhibit an universal
large-$p$ behaviour.

\section{Correlation function of gauge invariant observables}

An interesting  quantity to study is the two-point function
$\langle W(p,C)W^\dagger (p,C')\rangle$, where $W(p,C)$ has been
defined via Eqs.~\re{wline}, \re{linedef0}. It represents the
correlation function of two straight parallel Wilson lines of equal
length, each carrying a transverse momentum $p$.
In four dimensions such a correlator
was investigated by Gross {\em et al.} \cite{Gross}. They
compared the noncommutative two-point function of Wilson lines (for
large $|p|$ and in the 't Hooft limit $N\to \infty$, $g^2 N$ fixed)
with the Wilson loop in ordinary gauge  
theories in the  
Coulomb phase, which is a function of $T/L$ when the loop is a
rectangle of size $T\times L$. There, even though the expansion
parameter gets big for large enough $T$, an exponential form is obtained 
using perturbation theory by summing over ladder diagrams which
dominate when $T\gg L$. In four-dimensional noncommutative gauge
theories the length of the line $|p\th|$ plays the role of $T$ and the 
momentum $|p|$ plays the role of $1/L$, so that computing the
two-point function of the open Wilson lines at large momentum
corresponds to computing the expectation value of a rectangular Wilson 
loop in the commutative case. By resumming the ladder diagrams, the
correlator was found to grow exponentially at large momentum, corrections to the leading result having to be ascribed 
to finite $|p|$ and to subleading terms in  $N$. 
Actually, in four dimensions, at large momentum, the line correlator
was shown to provide the universal behaviour of the two-point function of any
noncommutative gauge invariant operator.

In two dimensions a similar situation occurs for planar diagrams, where
the dependence on $\th$ intervenes just through the length of the line $l$.

We quantize the theory Eq.~\re{act} in the light-cone gauge $A_-=0$ at {\em equal
times}, the free propagator having the following causal expression
(WML prescription)
\begin{equation}
\label{WMLprop}
D^{WML}_{++}(x)={1\over {2\pi}}\,\frac{x^{-}}{-x^{+}+i\epsilon x^{-}}\,,
\end{equation}
first proposed by T.T. Wu \cite{Wu}. In turn this propagator is nothing
but the restriction in two dimensions of the expression proposed 
by S. Mandelstam and G. Leibbrandt \cite{Leib} in four
dimensions~\footnote{In dimensions higher than two, where physical
degrees of freedom are switched on (transverse ``gluons''), this
causal prescription is the only acceptable one \cite{libro}.}  and 
derived by means of a canonical quantization in \cite{bosco}.

This form of the propagator allows a smooth transition to an Euclidean 
formulation, where momentum integrals are performed by means of a
``symmetric integration'' \cite{Wu}.
Were the theory quantized on the light-front,
the following free propagator would ensue
\begin{equation}
\label{hoop}
D^{l-f}_{++}(x)=-\frac{i}{2}|x^-|\delta(x^+),
\end{equation}
which cannot be smoothly continued  to Euclidean variables. Moreover, when 
combined with the Moyal $\star$-product, it gives rise to singularities which
cannot be cured \cite{torr}. Therefore the light-front formulation will not be 
further considered in the sequel.

The light-cone gauge gives rise to
other important features like the decoupling of Faddeev-Popov ghosts,
which occurs also in the noncommutative case \cite{Das}, and
the absence  of the triple gluon vertex in
two dimensions.
Consequently the computation of the Wilson line correlator
is enormously simplified.

With a forethought to the next section, with no loss of generality thanks to
the persisting boost invariance,
we choose the path
$C$ stretching along $x^0$, so that $p$ points in the $x^1$ direction
\footnote{A generic choice is discussed in Appendix B.}. In turn the
parallel path $C'$ is chosen at a distance $\Delta$ from $C$.
We begin by expanding the line operators
\beeq \label{pertline}
&&W(p,C)\,
=\,
\sum_{j=0}^\infty \,(ig)^j \int d^2x\int_{\z_j>\z_{j-1}>\ldots>\z_1}
[d\z]\, \tr
A(x+\z_1) \star  \ldots \star A(x+\z_j)\star e^{ipx}
\no\\
&&W^\dagger(p,C')\,
=\,
\sum_{j=0}^\infty \,(-ig)^j \int d^2x
\int_{\z'_j>\z'_{j-1}>\ldots>\z'_1} [d\z']\, \tr
A(x+\z'_j)  \star \ldots \star A(x+\z'_1)\star
e^{-ipx}\,.\no\\
&&
\eeeq
The variables $\z$ and $\z'$ can be conveniently parameterized as
$\z=(\s,-\Delta)$ and $\z'=(\s',0)$, with $\s=s\ l^0$, $0\le s\le 1$, $\Delta$ 
a constant quantity
in the $x^1$-direction and $l^0=-\theta p_1$, according to Eq.~(\ref{length}).
Eq.~(\ref{pertline}) can therefore be conveniently rewritten as
\beeq \label{nuopertline}
W(p,C)\,
&=&\,e^{ip\cdot \Delta}
\sum_{j=0}^\infty \,(ig)^j \int d^2x\int_{\s_j>\s_{j-1}>\ldots>\s_1}
[d\s]\, \tr
A(x+\s_1) \star  \ldots \star A(x+\s_j)\star e^{ipx}
\no\\
W^\dagger(p,C')\,
&=&\,
\sum_{j=0}^\infty \,(-ig)^j \int d^2x
\int_{\s'_j>\s'_{j-1}>\ldots>\s'_1} [d\s']\, \tr
A(x+\s'_j)  \star \ldots \star A(x+\s'_1)\star
e^{-ipx}.\no\\
&&
\eeeq

In the following the trivial phase factor $e^{ip\cdot \Delta}$ will be
omitted and the dependence on $C$, $C'$ will be understood.
We then contract the $A$'s in such a way that the resulting diagram
is of leading order in $N$, which yields, according to Eq.~(\ref{WMLprop}),
at a fixed perturbative
order $(g^2)^n$ 
\beq \label{1graph}
(-1)^{n-k}\lp\frac{N}{4\pi}\rp^n 
\int [d\s]\, [d\s']\int d^2x \,e^{ipx}\prod_{j=1}^k
\frac{x^0+f_j(\s,\s',\th p)
-x^1}{-x^0-f_j(\s,\s',\th p)-x^1}\,,\qquad n\ge 1,
\eeq
where $k$ is the number of propagators connecting the two lines~\footnote
{The contribution of propagators starting and ending on the same line
factorizes and amounts to $\lp -N/(4\pi) \rp^{n-k}$.}   and
$f_j(\s,\s',\th p)$ is a linear function of its variables
depending  on the topology;
the integration region for the $2n$ geometric
variables is understood and the phase factors containing the
noncommutativity parameter have been absorbed in the function
$f_j(\s,\s',\th p)$. 
We stress that, remarkably, factorization of propagators in coordinate
variables  occurs just in those diagrams which are dominant at
large $N$. This feature in turn makes $\th$-dependence trivial, as it
will be shortly cleared, since it intervenes just through the length
of the line $l$. A non-trivial example of a diagram which is leading
in $N$ and therefore ``planar'' in our terminology at $\OO (g^6)$ is
reported in Appendix~A.
Diagrams without any propagator connecting
the two lines will be disregarded. They provide indeed a contribution
concentrated at $p=0$, which is uninteresting for our purposes.

The procedure to be employed to select leading contributions in $N$
will be soon cleared out, but 
we now concentrate on the volume integration in
Eq.~\re{1graph}. Surprisingly enough, it will turn out that all
integrals will give the same result, no matter what the function
$f_j(\s,\s',\th p)$ is, \ie no matter what topology we choose in the
set of planar diagrams.
This finding is a direct consequence of the integration over the
world volume $d^2x$, required by noncommutative gauge invariance,
and of the orthogonality of the momentum with respect to the direction of
the open lines. In so doing the $\theta$-dependence is washed out
apart from its occurrence in $l^0$.

In order to provide a correct formulation of the theory, 
continuation to Euclidean variables is required: $x^0 \to i x_2$;
we recall that, to keep the basic algebra unchanged, the 
noncommutativity parameter $\theta$ has also
to be simultaneously continued to an imaginary value: $\theta \to i \theta$.

A symmetric integration \cite{Wu,Stau} 
then provides the natural regularization in Eq.~(\ref{1graph})
\beq\label{x1int}
\int d^2x \, e^{-ip\cdot x} \prod_{j=1}^k
\frac{x_1+i(f_j(\s,\s',\th p)+x_2)}{x_1-i(f_j(\s,\s',\th p)+x_2)}
=(-1)^k\frac{4\pi k}{p^2}.
\eeq
Hence the integration over the geometrical variables in Eq.~(\ref{1graph}) is
straightforwardly carried out and yields 
\beq \label{ordern}
\frac{4 \pi k}{p^2} \lp\frac{Nl^{2}}{4\pi}\rp^n
\frac1{n_1!\,n_2!}\,,
\eeq
where $l=|p\th|$ is the total length of the line and  $n_1$, $n_2$ are 
the number of legs stretching out of the first and the second line,
respectively ($n_1+n_2=2n$). 
(A detailed derivation of Eqs.~\re{x1int} and \re{ordern} is contained in 
Appendix~B.) 
As announced, Eq.~\re{ordern} displays a trivial dependence on the
topology of the graph, the only remnant being ${n_1!\,n_2!}$ in the
denominator. Thus, although resumming even only leading contributions
in $N$ may have seemed a formidable task when we started, it has now
become feasible, provided the exact number of different configurations 
with fixed $n_1$, $n_2$ is known.

One can see from Eq.~\re{pertline} that in the correlator $\langle
W(p)W^\dagger (p)\rangle$ the colour factors enter as 
$\tr \lq t^{A_1} t^{A_2} \ldots t^{A_{n_1}}\rq 
\tr \lq t^{B_1} t^{B_2} \ldots
t^{B_{n_2}}\rq $, where $A^j$, $B^m$ are indices of $U(N)$ and have to 
be contracted in pairs so that the result of the product of traces is
$(N/2)^n$. 
Let $j=1, \ldots, 2q$ be the indices to be contracted inside the first 
trace and $S_q$ the number of ways you can form  $q$ ``good'' pairs in
it. Imagine you have rearranged the order of matrices in that trace, so 
that the ones to be coupled  are adjacent and come in the first
$2q$ positions.
To avoid crossovers, which would reduce the power of $N$, the first
element can be paired only with $t^{2p}$, 
$p=1,\ldots,q$. In between
these two we are left with $2p-2$ matrices, and after $t^{2p}$ with $2q-2p$.
Therefore we are to face the same problem with a reduced number of
elements, namely we have to solve for $S_{p-1}$ and $S_{q-p}$. As a
consequence, a recursive relation can be established
\beq \label{recursion}
\left\{ \begin{array}{ll}
S_q=\sum_{p=1}^q S_{p-1}S_{q-p}&\qquad q\geq 1
\\
S_0=1&\qquad \mbox{no pairs}
\end{array}\right.
\eeq
which is solved by
\beq\label{solrec}
S_q=\frac{2^{2q}}{(q+1)!}\frac{\G(q+\half)}{\G(\half)}=
\left\{ \begin{array}{ll}
2^q\,\frac{(2q-1)!!}{(q+1)!}& \qquad q\geq 1\\
S_0=1\,.&
\end{array}\right.
\eeq
Analogously, the number of ways in which we can form $r$ pairs  on the
second line and maintain the leading power in $N$ is given by
$S_r$. (Obviously, the number of spare legs on each line has to match, 
\ie $r=\frac{n_2-n_1}2+q$.)

If $q,\,r\neq 0$, the product of traces reduces to
\beq 
\lp \frac{N}2\rp^{q+r}\tr \lq t^{A_1} t^{A_2} \ldots t^{A_{k}}\rq 
\tr \lq t^{B_1} t^{B_2} \ldots
t^{B_{k}}\rq \,, 
\eeq
where $k=n_1-2q=n_2-2r$. 
In addition matrices in one of the trace can be cycled (but not
swapped!),  leading to a
factor $k$ in the counting. Eventually, we have to take into account the
number of possibilities we had originally to choose a group of $k$
contiguous  matrices both from the first and the second trace, which
amounts to $n_1 \cdot n_2$.
At this stage the resummed contribution to the correlator of graphs with pairs
on both lines $\langle W(p)W^\dagger (p)\rangle_{\mathcal{P}\mathcal{P}}$
can be readily written 
\beeq
\label{pp}
&&\langle W(p)W^\dagger (p)\rangle_{\mathcal{P}\mathcal{P}}=
\frac{4\pi}{p^2} \sum_{n=1} 
\sum_{q=1}^{\lq\frac{n_1-1}2\rq}\sum_{r=1}^{\lq\frac{n_2-1}2\rq} (-1)^{q+r}
\tau^{2n} \,  k^2\cdot
n_1\cdot n_2 \cdot S_q \cdot S_r \cdot \frac1{n_1!} \cdot \frac1{n_2!} \\
&&\quad=
\frac{4\pi}{p^2} \sum_{r=1} \sum_{q=1} \sum_{k=1} (2i\tau)^{2q+2r} \tau^{2k}\,
\frac{k^2}{(q+1)!(r+1)!} \,\frac{\G(q+\half)}{(2q+k-1)!\G(\half)} \,
\frac{\G(r+\half)}{(2r+k-1)!\G(\half)} \,,\no
\eeeq
where $\tau=\sqrt{\frac{g^2 N l^2}{4\pi}}$.

Consider then the case  with $q=0$ and $r\neq 0$ (or viceversa). Following
the instructions provided in the previous example, it is easy to
build up the contribution $\langle W(p)W^\dagger
(p)\rangle_{\mathcal{P}}$ of graphs with pairs only on one line
(recall $k=n_1$)
\beeq
\label{p}
&&\langle W(p)W^\dagger (p)\rangle_{\mathcal{P}}=
\frac{8\pi}{p^2} \sum_{n=1} 
\sum_{r=1}^{\lq\frac{n_2-1}2\rq} (-1)^r
\tau^{2n} \, k\cdot 
n_1\cdot n_2 \cdot S_0 \cdot S_r \cdot \frac1{n_1!} \cdot \frac1{n_2!} =\\
&&\phantom{\langle W(p)W^\dagger (p)\rangle_{\mathcal{P}}}=
\frac{8\pi}{p^2}  \sum_{r=1} \sum_{k=1} (-1)^r 2^{2r} (\tau^2)^{r+k}\,
\frac{k}{(r+1)!} \,
\frac{\G(r+\half)}{(2r+k-1)!\G(\half)} \frac1{(k-1)!}\,.\no
\eeeq

The last piece entering the calculation of $\langle W(p)W^\dagger
(p)\rangle$ is due to graphs with no pairs at all ($q=r=0$, \ie
$n_1=n_2=n$), which simply reads 
\beq \label{ladder}
\langle W(p)W^\dagger (p)\rangle_{\mathcal{L}}=
\frac{4\pi}{p^2} \sum_{n=1} (\tau^2)^{n}\,
\frac1{[(n-1)!]^2} = \frac{4\pi \tau^2}{p^2} I_0(2\tau) \,.
\eeq
Notice that these graphs are ladder graphs in the sense of
\cite{Gross}, and only their multiplicity differs from the
four-dimensional case. In fact we saw here that a factor $n$ arises due to
the cyclicity property of traces. Thereafter, it is not a fortuitous 
coincidence that our result Eq.~\re{ladder} qualitatively agrees with
Eq.~(3.24)  of Gross {\it et al}.

Summing Eqs.~\re{pp}, \re{p}, \re{ladder}, after suitable
rearrangements, we get   
\beeq \label{complete}
&&\langle W(p)W^\dagger (p)\rangle=
\frac{4\pi \tau^2}{p^2}\Big\{
2\sum_{r=1} \frac{(-4)^r}{(r+1)!} \, 
\frac{\G(r+\half)}{\G(\half)} \lq \tau I_{2r+1} (2\tau)
+ I_{2r} (2\tau) \rq +\no \\
&&\phantom{\langle W(p)W^\dagger (p)\rangle=\frac{4\pi}{p^2}}
+  I_0(2\tau)+\sum_{r=1} r^2 \,  f_r^2 (\tau)  \Big\}\,,
\eeeq
with $I_q(2\tau)$ a modified Bessel function. The function  $f_r (\tau)$
can be written as 
\beq \label{fg}
f_r (\tau)=-\frac{1}{4\pi i}\int_{\gamma-i\infty}^{\gamma+i\infty}
\,dz\ e^{z\tau}\ z^{2-r}\left[1+\frac{2}{z^2}-\sqrt{1+\frac{4}{z^2}}\,\right],
\qquad \gamma> 0\ .
\eeq

Using an analogous Laplace representation for the Bessel functions,
all the sums in Eq.~(\ref{complete}) can be performed, leading to
\beeq \label{completata}
&&\langle W(p)W^\dagger (p)\rangle=\frac{4\pi \tau^2}{p^2}\lq
I_0(2\tau)+
\int_{\gamma-i\infty}^{\gamma+i\infty}\ \frac{dz}{8\pi i}\frac{z+\sqrt{z^2-4}}
{\sqrt{z^2-4}} e^{z\tau}\right. \\ \no
&& \times \left(\sqrt{1+(z-\sqrt{z^2-4})^2}-1-\frac12 (z-\sqrt{z^2-4})^2
\right)  \left(z+\sqrt{z^2-4}+2\tau \right)\\ \no
&&+
\int_{\nu_1-i\infty}^{\nu_1+i\infty} \int_{\nu_2-i\infty}^{\nu_2+i\infty}
\frac{dz\ dw}{(2\pi i)^2}  e^{(z+w)\tau} \frac{z^3w^3(1+zw)}{4(zw-1)^3}
\left(1+\frac{2}{z^2}-\sqrt{1+\frac{4}{z^2}}\right)\\ \no
&&\left. \times \left(1+\frac{2}{w^2}-\sqrt{1+\frac{4}{w^2}}\right)
\rq
\,,
\eeeq
where $\nu_1,\nu_2>1$ and $\gamma> 2 $.

One can realize that, at large $\tau$, the term with the double integration
dominates and $\langle W(p)W^\dagger (p)\rangle$ 
increases like $\exp (2\tau)=\exp(\sqrt{g^2 N l^2/\pi})$, disregarding a
(small) power correction.
As expected, the correlator depends on the 't Hooft coupling $\sqrt{g^2 N}$;
what is remarkable is that  its asymptotics is an exponential linearly 
increasing with the line momentum $|p|$. This is reminiscent of what was found 
by \cite{Gross} for its four-dimensional analog.
Nevertheless, whereas in four dimensions corrections to the leading
result have to be ascribed not only to subleading terms in  $N$ but
also 
to finite $|p|$, our result Eq.~\re{completata} is exact and receives
corrections only $\mathcal{O} (1/N)$.

In the next section we will turn our attention to a nonperturbative
derivation of the noncommutative Wilson lines correlator, and see how
it compares with Eq.~\re{completata} at large $|p|$. To this regard,
it is worth noticing 
that although Eq.~\re{completata} follows from a perturbative analysis,
having resummed all orders, it holds also at
large $g^2 N$. This fact will prove very useful in the sequel.

The situation here closely resembles the four dimensional case ($\NN
=4$ NCSYM) \cite{Gross}, where
the exponential growth in the line momentum of $\langle W(p)W^\dagger
(p)\rangle$ and its dependence on the 't Hooft coupling $\sqrt{ g^2N}$
allow to extrapolate from the perturbative regime to the strong
coupling regime. Many features of this structure are shared also by
the dual supergravity description of NCSYM (AdS/CFT
correspondence).
The parallelism between two-dimensional YM and ($\NN=4$ SYM) is
suggested also by the ordinary (\ie commutative) case. In fact, it was
shown in \cite{Zarembo} that perturbative computations of the expectation
value of a circular Wilson loop and a pair of infinite anti-parallel lines,
obtained by resumming an infinite class of ladder-like planar
diagrams, when extrapolated to strong coupling, produce an
expectation value peculiar of the results of AdS/CFT correspondence, 
namely $\langle W\rangle \sim \exp ({\rm const.}\sqrt{ g^2N})$. Though
resummation is a very complicated problem, in general, simplifications 
occur thanks to supersymmetry. One could be tempted to conjecture that 
the absence of transverse degrees of freedom and the 
finiteness of the theory in two dimensions play the role of
supersymmetry in dimension four.

\section{An approach based on the Morita equivalence}

When both coordinates are compactified to form a torus, 
a remarkable symmetry, called Morita equivalence \cite{Schwarz}, relates different
noncommutative gauge theories living on different noncommutative tori: the
duality group $SO(2,2,{\bf Z})$ has an $SL(2,{\bf Z})$ 
subgroup which acts as follows 
\begin{equation}
\label{morita1}
\left( \begin{array}{c}
m' \\
N'
\end{array} \right) =
\left( \begin{array}{cc}
a & b \\
c & d
\end{array} \right)
\left( \begin{array}{c}
m \\
N
\end{array} \right), \qquad \Theta'=\frac{c+d\Theta}{a+b\Theta},
\end{equation}
\begin{equation}
\label{morita2}
(R')^2=R^2 (a+b\Theta)^2,\qquad  (g')^2=g^2|a+b\Theta|, \qquad \tilde \Phi'=
(a+b\Theta)^2 \tilde \Phi-b(a+b\Theta),
\end{equation}
where $\Theta \equiv \theta/(2\pi R^2)$, $\tilde \Phi \equiv 2\pi R^2\Phi$, $\Phi$
being a background connection and $R$ the radius of the torus, which, for
simplicity, we assume to be square. The first entry $m$ denotes the
magnetic flux, while $N$ characterizes the gauge group $U(N)$. It is
not restrictive to consider the quantities $m$ and $\theta$ to be
positive. The parameters
of the transformation are integers, constrained by the condition
$ad-bc=1$.

The map in the equations above is flexible enough  to allow
for a commutative theory on the second torus by choosing $\Theta'=0$.
As a consequence, the parameter $d$ will be set equal to $-c/\Theta$,
$\Theta$  being a suitable rational quantity. 
In the sequel for notational convenience all the primed quantities
will acquire the subscript $c$ ($N'\equiv N_c,m'\equiv m_c,...$). 

We are eventually 
interested in a noncommutative
theory defined on a plane $(R \to \infty)$, with a trivial first Chern 
class ($m=0,\Phi=0$) and a gauge group $U(N)$ with a large $N$, since we want to
establish a comparison with the perturbative approach of the previous section.
Eqs.~\re{morita1}, \re{morita2} together
with the unimodularity condition allow us to eliminate the parameters
$a$, $b$ and $c$ in terms of ``physical'' quantities, namely $a=
\frac{N}{N_c}-\frac{\Theta m_c}{N}$, $b=\frac{m_c}{N}$ and 
$c=-\frac{\Theta N_c}{N}$. The quantity $d=\frac{N_c}{N}$ is  
a positive integer; as a consequence $N_c$ will also assume large values.
Moreover, it  follows that $m_c$ will be an integer multiple of $N$, $\Theta d$
will be integer and $\Theta d \frac{m_c}{N}-1$ will be an integer multiple
of $d$. All these constraints are compatible with suitable rational values
of $\Theta$.

Finally, the following relations between the radii of the tori
and between the coupling constants $$R_c=\frac{R}{d}, \qquad g_c^2=\frac{g^2}
{d}$$ will ensue. It is also useful to introduce the dimensionless
quantity $${\cal A}=4\pi^2 (g_c R_c)^2=4\pi^2 \frac{(gR)^2}{d^3},$$ related
to the total area of the commutative torus.

On the commutative torus we can develop the $U(N_c)$ Yang-Mills theory,
in particular compute its partition function, according to a
well established geometrical procedure \cite{Migdal-Rusakov}.

We start by considering the action
\beq
\label{action}
S=\frac{1}{4g_c^2}\int d^2 x \,  \tr\left[\left(F_{\mu\nu}-\frac{m_c}{2\pi R_c^2 N_c}
\epsilon_{\mu\nu} {\bf I}\right)\left(F^{\mu\nu}-\frac{m_c}{2\pi R_c^2 N_c}
\epsilon^{\mu\nu} {\bf I}\right)\right],
\eeq
where the explicit expression for the background connection 
$\Phi_c=-\frac{m_c}{2\pi R_c^2 N_c} {\bf I}$ has been introduced.

The quantity $m_c$ is just the first Chern class of the $U(N_c)$ field
$$m_c=\frac{1}{4\pi}\int d^2x \, \tr\left[F_{\mu\nu}\epsilon^{\mu\nu}\right];$$
the action thereby becomes
\beq
\label{action1}
S=\frac{1}{4g_c^2}\int d^2x \, \tr\left[F_{\mu\nu}F^{\mu\nu}\right]-
\frac{2\pi^2m_c^2}{{\cal A}N_c}.
\eeq

The general Migdal's formula for the partition function on a torus reads \cite{Migdal-Rusakov}
\beq
\label{migdal}
{\cal Z}=\sum_{{\cal R}} \exp\left[-\frac{{\cal A}}{2}C_2({\cal R})\right],
\eeq
$C_2$ being the second Casimir operator in the representation ${\cal R}$.
The sum runs over all the irreducible representations of the gauge group
$U(N_c)$, which can be labelled by a strongly decreasing sequence of
integers $n_1>n_2>...>n_{N_c}$, related to a Young tableau. In terms of such integers,
the Casimir operator reads
\beq
\label{casimir}
C_2({\cal R})=\sum_{i=1}^{N_c}\left(n_i-\frac{N_c-1}{2}\right)^2-\frac{N_c(N_c^2-1)}
{12}.
\eeq

The constant term only affects the overall normalization in our case and,
therefore, will be disregarded in the sequel. By exploiting the invariance 
under permutations, Eq.~(\ref{migdal}) can be written as 
\beq
\label{migdal1}
{\cal Z}=\frac{1}{N_c!}\sum_{n_1\neq n_2\neq...n_{N_c}} 
\exp\left[-\frac{{\cal A}}{2}\sum_{i=1}^{N_c}\left(n_i-\frac{N_c-1}{2}\right)^2\right].
\eeq

In order to fix the Chern class $m_c$ we have to factorize the partition function
according to the decomposition $U(N_c)=U(1)\times SU(N_c)/Z_{N_c}$. This can be
achieved introducing the integers $$\bar n_i=n_i-n_1,\qquad i=2,...,N_c.$$
Eq.~(\ref{migdal1}) becomes
\beeq
\label{migdal2}
&&{\cal Z}=\frac{1}{N_c!}\sum_{n_1=-\infty}^{+\infty}
\sum_{\bar n_i\neq \bar n_j\neq 0} 
\exp\left[-\frac{{\cal A}N_c}{2}\left( n_1 -\frac{N_c-1}{2}+\frac{1}{N_c}
\sum_{i=2}^{N_c}\bar n_i\right)^2\right] \times \\ \nonumber
&&\ph{{\cal Z}=\frac{1}{N_c!}\sum_{n_1=-\infty}^{+\infty}
\sum_{\bar n_i\neq \bar n_j\neq 0} }
\exp\left[-\frac{{\cal A}}{2}\left(\sum_{i=2}^{N_c}\bar n_i^2-\frac{1}{N_c}
\left(\sum_{i=2}^{N_c}\bar n_i\right)^2\right)\right].
\eeeq
Using standard techniques, Eq.~(\ref{migdal2}) can be rewritten as
\beeq
\label{migdal3}
{\cal Z}&=&\frac{1}{N_cN_c!}\sum_{l,k=0}^{N_c-1}\sum_{n_1=-\infty}^{+\infty}
\exp\left[-\frac{{\cal A}N_c}{2}\left( n_1 +\frac{l}{N_c}
\right)^2\right]\\ \nonumber
&\times&\sum_{\bar n_i\neq \bar n_j\neq 0} 
\exp\left[-\frac{{\cal A}}{2}\left(\sum_{i=2}^{N_c}\bar n_i^2-\frac{1}{N_c}
\left(\sum_{i=2}^{N_c}\bar n_i\right)^2\right)-2\pi ik\left(\frac{l}{N_c}+
\frac{N_c-1}{2}- \frac{1}{N_c}
\sum_{i=2}^{N_c}\bar n_i\right)\right].
\eeeq
After a Poisson resummation over $n_1$, the partition function becomes
\beq
\label{poisson}
{\cal Z}=\sqrt{\frac{2\pi}{{\cal A}N_c}}\sum_{k=0}^{N_c-1}\sum_{n_1=-\infty}^
{+\infty} \delta_{N_c}(n_1-k)\exp\left[-\frac{2\pi^2 n_1^2}{{\cal A}N_c}\right]
{\cal Z}_k,
\eeq
$\delta_N$ being the $N$-periodic delta-function and 
\beeq
&&\!\!\!{\cal Z}_k=\frac{1}{N_c !}\sum_{\bar n_i\neq \bar n_j\neq 0} 
\exp\left[-\frac{{\cal A}}{2}\left(\sum_{i=2}^{N_c}\bar n_i^2-\frac{1}{N_c}
\left(\sum_{i=2}^{N_c}\bar n_i\right)^2\right)-2\pi ik\left(
\frac{N_c-1}{2}- \frac{1}{N_c}
\sum_{i=2}^{N_c}\bar n_i\right)\right]\,,\no\\
&& \label{sector}
\eeeq
being the $SU(N_c)$ partition function in the $k$-th 't Hooft sector
\cite{flux}.  

By retaining only the term with $k=m_c$ and cancelling the $U(1)$ contribution
against the background connection, after restoring the full permutation symmetry
with respect to the integers $n_i$, we obtain the final expression
\beeq
\label{final}
{\cal Z}&=&\sqrt{\frac{2\pi}{{\cal A}N_c}}\frac{1}{N_c !}\sum_{n_i\neq n_j}
\exp\left[-\frac{{\cal A}}{2}\left(\sum_{i=1}^{N_c} n_i^2-\frac{1}{N_c}
\left(\sum_{i=1}^{N_c}n_i\right)^2\right)\right]\\ \nonumber
&\times&\int_0^{2\pi}\frac{d\alpha}{\sqrt {\pi}} \exp\left[-\left(\alpha -\frac{2\pi}{N_c}
\sum_{i=1}^{N_c}n_i\right)^2 -2\pi im_c\left(
\frac{N_c-1}{2}- \frac{1}{N_c}
\sum_{i=1}^{N_c}n_i\right)\right].
\eeeq

\smallskip

Now we turn our attention to the calculation of gauge invariant observable
quantities of the noncommutative theory; we are  interested here on the
correlation function of two straight parallel Wilson lines of equal length, at
a distance $\Delta$,
lying on the noncommutative torus without winding around it, each carrying
a transverse momentum $p$ \cite{Guralnik,Griguolo}. The noncommutative torus will eventually
be decompacted by sending its radius $R\to \infty$.

The operator for an open Wilson line  on the noncommutative plane has
been defined in Eq.~\re{linedef0}.
On the noncommutative torus, with the line $C$ stretching along $x_2$, we have the expression
\beq
\label{linet}
W(k, C)=\frac{1}{4\pi^2 R^2}\int_0^{2\pi R} d^2 x \, \Omega_\star 
[x,C]\star \exp(i k x_1/R
),
\eeq
where $k$ is the integer associated to the transverse momentum $p=\frac{k}{R}$.
The no-winding condition entails the constraint $\theta k<2\pi R^2$, namely
$l=p \theta<2\pi R$, $l$ being the total length of the straight line. 
The normalization of $W(k)$ is chosen so that $W(0)=1$.

Now we exploit again the Morita equivalence in order to map the open Wilson line
on the noncommutative torus on a closed Polyakov loop of the ordinary Yang-Mills theory
winding $k$ times around the commutative torus in the $x_2$ direction
\beeq
\label{map}
W(k)&=&W^{(k)},\\ \nonumber
W^{(k)}&=&\frac{1}{4\pi^2 R_c^2}\int_0^{2\pi R_c} d^2x \frac{1}{N_c}
\tr\left[\Omega^{(k)}
(x_1)\right].
\eeeq
The trace is to be taken in the fundamental representation of $U(N_c)$ and
$\Omega^{(k)}(x_1)$ is the holonomy of the closed path \cite{Ambjorn,Guralnik,Griguolo}. This
holonomy is to be computed in the flux sector $m_c$, singled out in the
decomposition $U(N_c)=U(1)\times SU(N_c)/Z_{N_c}$. Again the $U(1)$
contribution is taken in the trivial sector and cancels against the background
connection in the classical action.

The correlation function of two straight parallel open Wilson lines
$C$, $C'$, at a distance $\Delta$
in the $x_1$-direction reads \cite{Griguolo}
\beeq
\label{corre}
&&{\cal W}_2(k)\equiv <W(k,C)W(-k,C')>\\  \nonumber
&=&\ \exp\left(ip\cdot \Delta\right)\frac{1}{2\pi R_c} 
\int_0^{2\pi R_c} dx <\frac{1}{N_c} \tr\left[
\Omega^{(k)}(x)\right]
\frac{1}{N_c} \tr\left[\Omega^{(-k)}(0)\right]>,
\eeeq
where translational invariance in the $x_1$-direction has been taken into
account and thereby three trivial integrations have been performed. The
normalization in Eq.~(\ref{corre}) is such that ${\cal W}_2(0)=1$. Again
the trivial phase factor $\exp(ip\cdot \Delta)$ will be disregarded.

The integrand in Eq.~(\ref{corre}) can be computed by resorting to the
Migdal-Rusakov's formula for $U(N_c)$
\beeq
\label{mig-ru}
&&\frac{1}{N_c^2} <\tr\left[\Omega^{(k)}(x)\right] \tr\left[\Omega^{(-k)}(0)\right]>
=\frac{1}{{\cal Z}N_c^2}\sum_{{\cal R}{\cal S}}\exp\left[-\frac{{\cal A}}{2}\left(1-
\frac{x}{2\pi R_c}\right)C_2({\cal R})\right] \\ \nonumber
&&\times \exp\left[-\frac{{\cal A}}{2}\frac{x}{2\pi R_c}C_2({\cal S})\right]
\int dU_1 \chi_{{\cal R}}(U_1)\chi_{F}(U_1^{k})\chi^{\dagger}_{{\cal S}}(U_1)
\int dU_2 \chi_{{\cal S}}(U_2)\chi_{F}(U_2^{-k})\chi^{\dagger}_{{\cal R}}(U_2),
\eeeq
$\chi_{F}$ being the character of the fundamental representation of $U(N_c)$.

\smallskip

By repeating the procedure we have followed in computing the partition function,
keeping again the projection onto the $m_c$ sector in the decomposition $U(N_c)=U(1)
\times SU(N_c)/Z_{N_c}$ and subtracting the classical background, we get
\beeq
\label{correp}
&&\frac{1}{N_c^2} <\tr\left[\Omega^{(k)}(x)\right] \tr\left[\Omega^{(-k)}(0)\right]>
=\frac{1}{{\cal Z}N_c!}\sqrt{\frac{2\pi}{{\cal A}N_c}}\exp\left[
-\frac{k^2 x {\cal A}}{4\pi R_c}\left(1-\frac{x}{2\pi R_c N_c}\right)\right]
\no \\ \nonumber
&&\times\sum_{n_i\neq n_j}
\exp\left[-\frac{{\cal A}}{2}\left(\sum_{i=1}^{N_c} n_i^2-\frac{1}{N_c}
\left(\sum_{i=1}^{N_c}n_i\right)^2\right)\right]\frac{1}{N_c}
\sum_{j=1}^{N_c}\exp\left[-\frac{xk{\cal A}}{2\pi R_c}\left(n_j-
\frac{1}{N_c}\sum_{i=1}^{N_c}n_i\right)\right] \\ 
&&\times\int_0^{2\pi}\frac{d\alpha}{\sqrt {\pi}} \exp\left[-\left(\alpha -\frac{2\pi}{N_c}
\sum_{i=1}^{N_c}n_i\right)^2 -2\pi im_c\left(
\frac{N_c-1}{2}- \frac{1}{N_c}
\sum_{i=1}^{N_c}n_i\right)\right].
\eeeq

\section{The planar phase}

Eqs.~\re{final}, \re{correp} entail $N_c$ sums over the different integers
$n_i$ which can take any value between $-\infty$ and $+\infty$. To this regard
they might be interpreted as mimicking a fermionic model, the $n_i$ labelling
the energetic levels which can be either empty or singly occupied. As a
consequence of the $SU(N_c)/Z_{N_c}$ symmetry, those equations are manifestly
invariant under a simultaneous shift of all the $n_i$ by an integer. 

Obviously, plenty of different configurations are possible and to sum
over all of them is beyond  reach. We are therefore seeking for 
configurations which may be dominant in particular physical regimes.
In a recent paper \cite{Mandal}, three basic different regimes have been presented
for a scalar noncommutative theory in two dimensions, when approximated
by means of a $M\times M$ matrix model. Three different phases (disordered,
planar and GMS \cite{Gopakumar}) are possible according to the behaviour of the
noncommutativity parameter $\theta$ with respect to the integer $M$
which is to be sent  
eventually to $\infty$ ($\theta \sim M^{\nu}$ with $\nu <1$, $\n=1$, $\n>1$, 
respectively).
This integer in turn is related to a large distance cutoff $L$ of the theory,
which can be identified with the length of the side of a  square torus.
 
In our case we are facing  a gauge theory: we have previously obtained a
perturbative expression for the two Wilson line correlator on the
noncommutative plane. In order to justify keeping only planar diagrams, we have
to resort to a large-$N$ approximation, $U(N)$ being the relevant gauge group.

It is not granted {\it a priori} that a comparison between perturbative and
non-perturbative results may be possible at all. The common wisdom is that a comparison
(if any) may concern the planar phase; this was actually the finding for 
${\cal N}=4$ NCSYM in four dimensions, via AdS/CFT correspondence \cite{Gross}.

In two dimensions  as well dramatic simplifications are expected to
take place, thanks to  
the absence of transverse degrees of freedom and to the onset of the invariance
under area-preserving diffeomorphisms. Then the Morita equivalence should
provide the required bridge.

On the non-perturbative side some constraints may be imposed; coming back 
to the quantities introduced
in the previous section to parameterize the Morita transformation,
we start by considering large $N$ values in order to comply
with our perturbative treatment of Sect.~3; this forces even larger
values for $N_c$, the ratio $d$ being a positive integer,
and a large value for the ``commutative'' flux $m_c$. 

Eqs.~\re{final}, \re{correp} exhibit a Gaussian damping with
respect to the ``occupation numbers'' $n_i$, which
suggests that a kind of saddle-point approximation may be feasible. 
The most favoured configurations
are those with minimal fluctuations. This
happens when the integers $n_i$  assume adjacent values.

Thanks to the invariance under a global translation of the $n_i$'s and
choosing for the sake of simplicity an odd value for $N_c$, we can set $\sum_{i=1}^{N_c}
n_i=0$; the other equivalent configurations will be disregarded as
they will eventually cancel in normalized quantities like correlators.
 Eq.~(\ref{final}) then becomes
\beq
\label{finalapp}
{\cal Z}\simeq\sqrt{\frac{2\pi}{{\cal A}N_c}}\frac{1}{N_c !}
\exp\left[-\frac{{\cal A}}{2}\frac{(N_c-1)N_c(N_c+1)}{12}\right]
\int_0^{2\pi}\frac{d\alpha}{\sqrt {\pi}} \exp(-\alpha^2).
\eeq
Other sets of $n_i$ lead to an extra suppression factor behaving at
least as $\exp(-{\cal A}N_c)$.

Now we turn our attention to the two line correlator, namely to
Eq.~(\ref{correp}). 
In order to retain the damping character of the overall exponential
factor, we must require that $|k|$ is not increasing faster
than $N_c^2$. 
However, the configurations with the $n_i$'s assuming adjacent values
remain the most relevant ones for the correlator only provided  $|k|$
is not increasing faster than $N_c$. As a matter of fact, in
Eq.~\re{correp} different configurations would produce competing
exponential corrections behaving as $\exp (-{\cal A}N_c \de)$ and 
$\exp (|k| {\cal A} \de)$, $\de$ being a typical gap in the
distribution of the  $n_i$'s.
Then, if $|k|<N_c$, Eq.~(\ref{correp}) can be approximated  
following the same procedure
we applied to derive Eq.~(\ref{finalapp}), leading to
\beeq
\label{correp1}
&&\frac{1}{N_c^2} <\tr\left[\Omega^{(k)}(x)\right] \tr\left[\Omega^{(-k)}(0)\right]>
\simeq \frac{1}{{\cal Z}N_c!}\sqrt{\frac{2\pi}{{\cal A}N_c}}\exp\left[
-\frac{k^2 x {\cal A}}{4\pi R_c}\left(1-\frac{x}{2\pi R_c N_c}\right)\right]
 \nonumber \\ 
&& \times \exp\left[-\frac{{\cal A}(N_c-1)N_c(N_c+1)}{24}\right]\frac{1}{N_c}
\frac{\sinh \frac{x|k|{\cal A}N_c}{4\pi R_c}}
{\sinh \frac{x|k|\cal{A}}{4\pi R_c}}
\int_0^{2\pi}\frac{d\alpha}{\sqrt {\pi}} \exp(-\alpha^2).
\eeeq

Taking Eq.~(\ref{finalapp}) into account, we eventually get, for large values of $N_c$,
\beq
\label{correap1}
\frac{1}{N_c^2} <\tr\left[\Omega^{(k)}(x)\right]
\tr\left[\Omega^{(-k)}(0)\right]> \simeq 
\frac{1}{N_c}\exp\left[\frac{x{\cal A}}{4\pi R_c}|k|(N_c-1-|k|)\right]\,.
\eeq

Introducing Eq.~(\ref{correap1}) in Eq.~(\ref{corre}) and performing the
integration over $x$, we obtain
\beq
\label{corref1}
{\cal W}_2(k)\equiv <W(k)W(-k)>\simeq
\frac{\exp\left[\frac{{\cal A}}{2}|k|(N_c-1-|k|)\right]-1}
{\frac{{\cal A}}{2}N_c|k|(N_c-1-|k|)}. 
\eeq

Remarkably, when $|k|<N_c-1$, we find a correlation function exponentially
{\it increasing} with $|k|$, in qualitative agreement with an analogous finding 
in \cite{Gross} and with our perturbative result in Sect.~3
\beq
\label{corref}
{\cal W}_2(k)\equiv <W(k)W(-k)>\simeq
\frac{\exp\left(\frac{{\cal A}}{2}|k|N_c\right)}
{\frac{{\cal A}}{2}|k|N_c^2}. 
\eeq

In order to reach a quantitative agreement we can consider a fine tuning
of the exponents. When comparing Eq.~(\ref{correp}) in the large-$N$
approximation 
with Eq.~(\ref{completata}), one might argue that their normalizations are
different; actually Eq.~(\ref{completata}) could also be normalized to be
unity at $p=0$; however this would only entail a trivial power factor.

If we trade the quantities ${\cal A}$ and $N_c$ for $\th$, $R$ and
$|c|$, by
means of their expressions (see the previous section) 
\beq
\label{trade}
{\cal A}=\frac{g^2\theta^3}
{2\pi R^4 |c|^3}, \qquad N_c= \frac{1}{\theta}2\pi R^2 N |c|,
\eeq
we get
\beq
\label{nuocorref}
{\cal W}_2(k)\simeq
\frac{\exp\left(p\frac{g^2N\theta^2}{2R|c|^2}\right)}
{p\frac{g^2N\theta^2}{2R|c|^2}}. 
\eeq
Agreement with the asymptotic behaviour of Eq.~(\ref{completata}) suggests
\beq
\label{tuning}
\theta \sim \frac{2R c^2}{g\sqrt{N\pi}} \,.
\eeq

Let us summarize at this point the various requirements to be met for
the consistency of our procedure. The radius
$R$ of the noncommutative torus has to be sent to $\infty$, since
we want to reach eventually the noncommutative plane, as well as $N$
of $U(N)$, so that planarity is entailed.
The validity of the saddle-point approximation we have hitherto
considered requires 
$${\cal A}N_c \sim c^2 \to \infty \,.$$ 
On the other hand, the no-winding condition $p\theta<2\pi R$ we have imposed in our non-perturbative
approach entails that $\theta$ cannot increase faster than $R$ when we consider
non-vanishing values of the transverse momentum. Hence, from
Eq.~\re{tuning}, it follows
$$ g\sqrt{N}\gtrsim c^2 p\,,$$
which implies that, for finite $p$, the 't Hooft coupling should be
large in this limit. 

In truth, the occurrence of a strong ('t Hooft) coupling regime is not 
surprising. We already noticed in Sect.~3 how the result
Eq.~\re{completata}, though following from a perturbative analysis,
holds also at large values of $g^2 N$. 
The situation here closely resembles the four dimensional case ($\NN
=4$ NCSYM) \cite{Gross}, in the fact that 
the exponential growth in the line momentum of $\langle W(p)W^\dagger
(p)\rangle$ and its dependence on the 't Hooft coupling $\sqrt{ g^2N}$
allow to extrapolate from the perturbative regime to the strong
coupling regime. 

If we remember that $R$ is the
natural cutoff of our formulation, the condition $\theta \sim R^{\nu}$ with
$\nu \leq 1$ is reminiscent of the analogous condition in \cite{Mandal} with respect
to the cutoff $M$ (or the torus side length $L$), 
related to  the dimension of their matrix model. 
Actually $\nu <1$ describes the disordered phase, where quantum effects are
dominant, while $\nu =1$ is a border-line value, related to the so-called
planar phase. The values $\nu >1$ would correspond to the GMS phase which
is inaccessible to our treatment. 

As a final remark we notice that Eq.~\re{corref1} changes dramatically 
when $|k| > N_c-1$, strongly deviating from the perturbative result
and thus possibly suggesting the onset of a new phase. Nevertheless,
we ought to recall that in this region the saddle-point approximation
we adopted no longer holds.

\section{Conclusions}

Two quite different approaches were pursued to calculate  the
correlator of two parallel Wilson lines in two space-time
dimensions. In the first one we
considered a perturbative expansion and were able to resum all planar
diagrams,
which is justified as far as the large-$N$ limit is concerned.
For planar diagrams, $\theta$-dependent phases resulting from 
noncommutativity play no role. This is indeed a particular phenomenon occurring
only in two dimensions, where transverse degrees of freedom are absent and the
theory exhibits invariance under area-preserving diffeomorphisms.

In spite of these  peculiarities, we still find a correlation function which 
increases exponentially with respect to the momentum associated to the lines,
in analogy with the result for ${\cal N}=4$ NCSYM in four dimensions found in
\cite{Gross} in the planar regime. We conjecture that the simplifying role
of supersymmetry is replaced in two dimensions by the triviality of
the phase space.

On the non-perturbative side in four dimensions a comparison with supergravity
results was possible via AdS/CFT correspondence.
In two dimensions we  exploited the Morita equivalence in order to map the 
open lines on the noncommutative torus (which eventually gets decompacted) into
two closed Wilson loops winding around the dual commutative torus. In turn, in a
commutative setting, the correlator can be obtained by well known geometrical 
techniques; this provides a non-perturbative solution and open the possibility
of a comparison with the perturbative result, provided they share a kinematical
region of validity. The common wisdom is that a comparison 
may be possible when  the theory is in the planar phase;
this was indeed confirmed for ${\cal N}=4$ NCSYM in four dimensions via AdS/CFT
correspondence.

In two dimensions planarity allows us to perform a saddle-point approximation
on the general Morita expression for the correlator, which nicely compares
with the perturbative result, exhibiting an exponential increase with respect
to the momentum $p$ in the region $k\lesssim N_c$. 
Beyond such a value, the approximation we used can no longer be
trusted, possibly calling for a new regime.
To clarify and properly understand this issue looks exciting and
promises further interesting insights into the theory.

\acknowledgments
Discussions with G. De Pol and L. Griguolo are gratefully acknowledged. One
of us (F.V.) is pleased to thank  K. Zarembo for helpful comments.

\section{Appendix A}
\begin{figure}[h]
\begin{center}
\epsfxsize=7cm
\epsffile{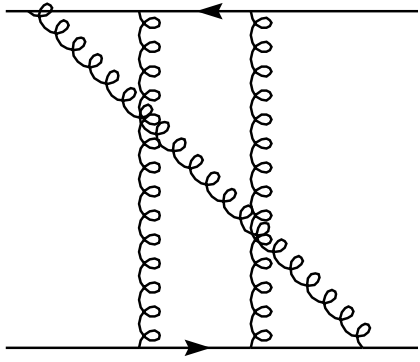}
\caption{A non-trivial leading diagram in the large-$N$ limit.}
\end{center}
\end{figure}

The diagram in Fig.1 entails the following integral in momentum space
\beeq\label{iex}
&&I=\int \frac{d^2 p_1}{[p_{1-}]^2} \, \frac{d^2 p_2}{[p_{2-}]^2} \, \frac{d^2
p_3}{[p_{3-}]^2} 
\, \de^{(2)} (p_1+p_2+p_3 -p) \\
&&\quad \quad\times
\exp \lq -ip_1(\z_1-\z'_3) -ip_2(\z_2-\z'_1)
-ip_3(\z_3-\z'_2)+i (p_1\th p_2+p_1 \th p_3)\rq, \no
\eeeq
where $p_-\equiv p_x-ip_y$ in Euclidean components. An easy calculation leads to
\beeq\label{propfact}
&&I=-\frac{\pi}{4} \int d^2x \, e^{-ipx}
\frac{x_1+i(x_2-\z_2+\z'_1)}{x_1-i(x_2-\z_2+\z'_1)} \no\\
&&\ph{I=\frac1{(2\pi)^2 \pi^2} \int }
\times \frac{x_1+i(x_2-\z_3+\z'_2)}{x_1-i(x_2-\z_3+\z'_2)}\, \times
\frac{x_1+i(x_2-\z_1+\z'_3-\th p)}{x_1-i(x_2-\z_1+\z'_3-\th p)},
\eeeq
which has precisely the form of the left-hand side of Eq.~(\ref{x1int}). 
Integration over the $\z$'s
has to be subsequently performed.
\section{Appendix B}
We start by proving Eq.~\re{x1int}.
Let us define
\beq \label{in}
I_k=\int d^2x \,e^{-ipx}\prod_{j=1}^k
\frac{x_1+F_{1,j}(\z,\z',\th p)
+i(F_{2,j}(\z,\z',\th p)+x_2)}
{x_1+F_{1,j}(\z,\z',\th p)-i(F_{2,j}(\z,\z',\th p)+x_2)}\,,
\eeq
where $F_j(\z,\z',\th p)=F_{1,j}(\z,\z',\th p)+iF_{2,j}(\z,\z',\th p)$ are 
linear functions of their variables
according to the topology of the diagram. 
This is a more general version of the integral
in Eq.~\re{x1int}, since we allow the parallel lines, parametrized
by $\z$, $\z'$,  to form an arbitrary angle 
with the frame and thus the functions $F_j$ become complex.
By means of a recursive argument the product of denominators in Eq.~\re{in}
can be easily simplified as
\beq \label{denos}
\prod_{j=1}^k
\frac1{x_1+F_j^*(\z,\z',\th p)-ix_2}=\sum_{m=1}^k \, \prod_{j\neq
m}\frac1{F_j^*-F_m^*}\,\frac1{x_1+F_m^*-ix_2}\,.
\eeq
Inserting Eq.~\re{denos} in Eq.~\re{in} and perfoming a shift in the
integration variables such that $x_1 \mapsto x_1 + F_{1,m}$, $x_2
\mapsto x_2 + F_{2,m}$, we get
\beeq\label{insimp}
&&I_k= \sum_{m=1}^k 
\, \prod_{j\neq m}\frac1{F_j^*-F_m^*}
\int d^2x \,e^{-ipx}\, \frac{\prod_{l=1}^k (x_1+F_l-F_m +ix_2)}{x_1-ix_2}\no \\
&& \ph{I_n}=
\sum_{m=1}^k  \sum_{q=0}^{k-1}  
\sum_{j_1<j_2 <\ldots <j_q }
(F_{j_1}-F_m) (F_{j_2}-F_m) \ldots (F_{j_q}-F_m)  \no \\
&& \ph{I_n}\times
\prod_{j\neq m}\frac1{F_j^*-F_m^*}
\int d^2x \,e^{-ipx}\, \frac{(x_1+ix_2)^{k-q}}{x_1-ix_2}
\,,
\eeeq
where the sum is over all possible choices of indices
$\{j_1,j_2,\ldots,j_q \}$ among 1 and $k$ (notice that it vanishes for
any $j_i=m$). 
We will demonstrate right ahead that the contributions to Eq.~\re{insimp} due  
to $q={0,1,\ldots, k-2}$ vanish and only the one due to 
$q=k-1$ survives, providing
\beq\label{ordern-1}
I_k= \sum_{m=1}^k  \, \prod_{j\neq m} 
\frac{F_{j}-F_m}{F_j^*-F_m^*} 
\int d^2x \,e^{-ipx}\, \frac{ x_1+ix_2}{x_1-ix_2}=
\frac{4\pi}{p^2}\, k\,(-e^{2i\psi})^k\,.
\eeq
The previous result can be reconstructed by recalling that $F_j$ is 
aligned with $l$ (see Eq.~(\ref{endpoints})). 
If we use a two-component vector notation and parametrize  
$\vec p=p(\cos \psi, \sin\psi)$, 
then $\vec F_j=|F_j| (\sin \psi, -\cos \psi)$, $\forall j$, so that 
Eq.~\re{ordern-1} ensues. In order to prove Eq.~\re{ordern}, we remark that,
in the case of
an arbitrary angle $\psi$ between the lines and the reference frame,
the factor $(-1)^{n-k}$ in Eq.~\re{1graph}, due to integrating over
propagators starting and ending on the same line, is replaced by
$(-e^{2i\psi})^{n-k}$.
Eventually, when performing the integration over the $2n$
line variables $\z^+$, $\z'^+$, Eq.~\re{ordern} is recovered, since
the factor $(-e^{2i\psi})^n$ cancels 
against its counterpart $(-e^{-2i\psi})^n$ originating from the 
$\z^+$ integrations. Notice that
Eq.~\re{x1int} corresponds to the choice 
$\psi=0$.

To conclude our proof we have to show the following 
\beq \label{canc}
\sum_{m=1}^k  \,
\sum_{j_1< j_2 < \ldots < j_q }
(F_{j_1}-F_m) (F_{j_2}-F_m) \ldots (F_{j_q}-F_m)
\prod_{j\neq m}\frac1{F_j^*-F_m^*} =0\,,
\eeq
for  $q={0,1,\ldots, k-2}$.
We start by rewriting the l.h.s. of Eq.~\re{canc} as
\beeq
\label{cancn-q}
&& (-e^{2i\psi})^q\sum_{m=1}^k  \,
\sum_{i_1< i_2 < \ldots < i_{k-q-1}\neq m }\frac1{F_{i_1}^*-F_m^*} 
\frac1{F_{i_2}^*-F_m^*} \ldots \frac1{F_{i_{n-q-1}}^*-F_m^*} =\no\\
&&(-e^{2i\psi})^q\sum_{m=1}^k  \, \sum_{i_1< i_2 < \ldots < i_{k-q-1}\neq m}
\frac{
\D_{k-q-1}(F_{i_1}^*,F_{i_2}^*, \ldots, F_{i_{k-q-1}}^*)
}{
\D_{k-q}(F_{i_1}^*,F_{i_2}^*, \ldots, F_{i_{k-q-1}}^*,F_m^* ) 
}\,,
\eeeq
where $\D_{l}$ is  the Vandermonde determinant of $l$ elements. The
sum is over all indices in $\{1,\ldots,k\}$ left over by the set
$\{j_1,\ldots,j_q,m\}$, which have been labelled   $\{i_1,\ldots,i_{k-q-1}\}$. 
In Eq.~\re{cancn-q} the index $m$ is not ordered with respect to the
$\{i_l\}$'s. By exploiting the symmetry properties of the Vandermonde
determinant in the denominator and relabelling $m=i_{k-q}$, we get 
(up to an  overall phase)
\beeq
\label{cancord}
&&\sum_{i_1< i_2 < \ldots < i_{k-q}} \D_{k-q}^{-1} \,
(F_{i_1}^*,F_{i_2}^*,\ldots, F_{i_{k-q-1}}^*,F_{i_{k-q}}^* ) \times \no \\
&&\sum_{r=1}^{k-q} (-1)^{k-q-r}
\D_{k-q-1}(F_{i_1}^*,\ldots,F_{i_{r-1}}^*, F_{i_{r+1}}^*, 
\ldots, F_{i_{k-q-1}}^*,F_{i_{k-q}}^*)\,,
\eeeq
which vanishes thanks to a well known property of determinants.

We conclude that Eq.~\re{x1int} indeed holds.

\end{document}